\documentclass[aps,preprint,groupedaddress]{revtex4}
\usepackage{graphicx}

\begin{document}

\title{Landau-Zener quantum tunneling in disordered nanomagnets}


\author{V.G.~Benza}
\affiliation{Dipartimento di Fisica e Matematica, Universita' dell'Insubria,
Como, I.N.F.M., sezione di Como, Italy}
\author{C.M.~Canali}
\affiliation{Division of Physics,
Department of Chemistry and Biomedical Sciences, 
Kalmar University, 391 82 Kalmar,
Sweden}
\author{G. Strini}
\affiliation{Dipartimento di Fisica, Universita' di Milano, Milano, Italy}


\date{\today}

\begin{abstract}

We study Landau-Zener macroscopic quantum transitions 
in ferromagnetic metal nanoparticles containing on the order of 100 atoms.
The model that we consider is described by an effective giant-spin
Hamiltonian, with a coupling to
a random transverse
magnetic field mimicking the effect of
quasiparticle excitations and structural disorder on the gap structure 
of the spin collective modes.
We find different types of time
evolutions depending on the interplay between the disorder 
in the transverse field
and the initial conditions of the system. 
In the absence of disorder, 
if the system starts from a low-energy state, there is one main
coherent quantum tunneling event where
the initial-state amplitude is completely depleted in favor of 
a few discrete states,
with nearby spin quantum numbers;
when starting from the highest excited
state, we observe complete inversion of the magnetization
through a peculiar ``backward cascade
evolution''.
In the random case, the disorder-averaged
transition probability
for a low-energy initial state becomes a smooth
distribution, which is nevertheless still sharply peaked 
around one of the transitions present in
the disorder-free case. On the other hand, the coherent backward cascade 
phenomenon
turns into a damped cascade with 
frustrated magnetic inversion.

\end{abstract}

\maketitle


\section{Introduction}
\label{intro}
Ferromagnetic transition metal 
nanoparticles\cite{billas1994, lederman1994, kodama1999}
and molecular nanomagnets \cite{sessoli1993,wernsdorfer2001} have
been actively studied over the past decade and are presently the subject
of strong interest and intense investigation. 
So far, interest in
ferromagnetic transition metal
nanoparticles has been  mainly motivated by their
relevance to high-performance information storage technology and spin 
electronics\cite{majetich1999, murray_science2000}. 
Recently a lot of progress
has been made in characterizing the physical properties of individual
ferromagnetic nanoparticles, 
such as their magnetic anisotropy\cite{jamet2001}.
However, reproducible and controlled fabrication is still difficult
and the understanding of their classical dynamics is still
not a fully solved problem\cite{garg0012157}.
Molecular magnets, 
on the other hand, are relatively simple and well 
characterized magnetic systems 
that offer 
the possibility of studying a rich interplay
of classical and {\it quantum} magnetic phenomena\cite{wernsdorfer2001}. 
Among the latter, the coherent
quantum tunneling of the 
magnetization in molecular magnets is one of the 
most fascinating phenomena\cite{qtm94,chudnovsky_tejada98}
\footnote{In
discussing quantum tunneling phenomena involving macroscopic variables 
it is important
to distinguish between {\it incoherent tunneling} of the system 
out of a classically metastable
potential well and  {\it coherent tunneling} between two classically degenerate
minima separated by an impenetrable barrier\cite{qtm94}. 
In the second case, coherent tunneling removes the degeneracy of the
two original ground states causing a level splitting. In this paper we will
loosely use the expression macroscopic quantum tunneling to indicate
macroscopic quantum coherence, that is coherent tunneling.}.
Thermally activated
quantum tunneling of the magnetization (QTM) has been observed experimentally
in molecular magnets such as Mn$_{12}$\cite{thomas1996,friedman1996} 
and Fe$_8$\cite{sangregorio1997}. In these experiments, 
a sequence of discrete steps in the magnetic hysteresis curve provides
a direct evidence of resonant coherent quantum tunneling between 
collective spin quantum states.
For Fe$_8$, there is strong evidence that at low temperatures
(below 360 mK) the system enters a quantum regime where the
reversal of the magnetization
is caused by a pure 
tunneling mechanism\cite{sangregorio1997,wernsdorfer1999_science}.
The occurrence of macroscopic
QTM has also been investigated in ferromagnetic nanoparticles.
Despite the effective spin of a few nanometer particle is typically
several orders of magnitude larger than the spin of a molecular magnet
(where $S=10$) earlier theoretical work\cite{qtm94} 
predicts that QTM in these systems
should be possible by applying an external field close to the
classical switching field in the direction opposite to the magnetization.
On the experimental front, some evidence of quantum effects was found
in switching field measurements in 
ferrimagnetic BaFeO nanoparticles\cite{wernsdorfer97}.
However, it is fair to say that experimental proof for QTM in 
single-domain nanoparticles is still a controversial issue.
Unambiguous evidence of
QTM can only be provided by the observation of level quantization of the
collective spin states like in the case of molecular magnets. 
For BaFeO
nanoparticles with $S = 10^5$,
the magnetic field steps associated with such quantizations
should be of the order of a $\Delta H=H_a/2S \approx 0.002$ mT (where
$H_a$ is the anisotropy field), which is 
too small even for the most sensitive and sophisticated magnetic 
measurement techniques, such as the  
new microSQUID set-up used in Ref.~\onlinecite{jamet2001}.
It seems clear that new experiments in this direction, presently underway,
should focus on nanoparticles containing on the order of a few hundred atoms.
In discussing ferromagnetic nanoparticles it is important to draw
a clear distinction between insulating particles, for which
the only low-energy degree of freedom is
the collective spin-orientation, and metal nanoparticles,
which have discrete particle-hole excitations in addition.
The common practice of modeling a magnetic particle by a spin Hamiltonian,
completely misses this aspect of metal physics. In general for  nanoparticles
containing a few thousand atoms, both types of
excitations are present in the low-energy quantum spectrum of transition metal
ferromagnetic nanoparticles. 

Recently the low-energy quantum states 
of individual
ferromagnetic metal nanoparticles have been directly probed by means of
single-electron-transistor (SET) 
spectroscopy\cite{gueron1999,deshmukh2001,mandar_thesis2002}. 
These experiments have 
indeed demonstrated the existence of a complex pattern of excitation spectra
and spurred the elaboration of 
adequate theoretical 
models\cite{cmc_ahm2000prl,ahm_cmc2001ssc,kleff2001prb,cmc_ac_ahm2002pap4} 
that include particle-hole and collective excitations on the same footing.
It is clear that the quasiparticle states will change when the collective
magnetization orientation is manipulated with an external field. Thus
itinerant quasiparticle excitations
can give rise to dissipation in the 
dynamics of the collective magnetization\cite{tatara1994}. 
These important and interesting features represent a considerable
complication that cannot be avoided in a 
theoretical treatment of macroscopic QTM
in ferromagnetic metal nanoparticles. Dissipation from particle-hole
excitations might be one of the reasons 
why preliminary experiments\cite{wernsdorfer2001} on
3 nm Fe nanoparticles with $S\approx 800$ yield broad
switching-field distribution widths that completely smear the expected
field-separation steps coming from level quantization.  
There is, however, a regime where the quantum description of small
transition metal clusters 
simplifies considerably, and their low-energy physics can be described by an
effective Hamiltonian with a single giant spin degree of freedom, like
that of a molecular magnet. Indeed, in Ref.~\onlinecite{ccmPRL03} it was argued
that a transition metal nanoparticle will behave like a molecular magnet
when the energy scale associated with the collective magnetization,
the magnetic anisotropy, is smaller than the typical energy scale associated
with the quasiparticle degree of freedom, $\delta$. This is the case
for transition metal nanoparticles when the number of atoms is on the order
of 100. 

In this paper we study Landau-Zener macroscopic QTM in transition
metal nanoparticles,
in the regime where they behave like molecular magnets. 
In Ref.~\onlinecite{ccmPRL03} it was shown that in this case 
the total spin of the
effective Hamiltonian describing the nanoparticle is specified by a Berry
curvature Chern number that characterizes the topologically nontrivial
dependence of the many-electron wavefunction on magnetization orientation.
A prescription was given to derive microscopically 
the effective spin Hamiltonian
by integrating out the quasiparticle degrees of freedom in the quantum
action constructed within an approximate spin-density-functional 
theory framework.
Here, however, we take a more pragmatic
view, and we assume that the effect of the quasiparticle degrees of freedom, 
when integrated
out, is to reshuffle randomly the 
pattern of avoided crossing gaps 
in the excitation spectrum of the collective-magnetization orientation
degree of freedom used to describe
uniaxial molecular magnets.
Although this procedure might appear {\it ad hoc}, we believe
that such a ``random matrix theory'' should capture part of the complicated
interleaved excitation spectra in real systems, including those
features due to surface-imperfections-induced randomness
that are likely to be important in these
very small grains. 
Our goal is to examine the effect of this randomness
on the coherent macroscopic quantum transitions
triggered by a magnetic field swept in the direction of the magnetization.
We neglect in the present analysis every effect arising from decoherence and
dissipation.
We find different time evolutions depending 
on the presence or absence of disorder {\it and}
on the initial conditions of the system.
In the absence of disorder,
if the initial state is close to the ground state, the system undergoes
a few coherent transitions, corresponding to macroscopic tunneling
of the collective spin between quasi-degenerate states. 
The asymptotic transition probability displays a small number
of discrete peaks, with one dominant contribution. In general we find
that the shift of the magnetization associated with the dominant transition is
not large but still significant.
A spectacular effect takes place if the system is prepared initially in a
state of high-energy. In this case, during its time evolution, the system
displays complete magnetization inversion, through a peculiar phenomenon that
we call ``backward cascade''. 
When disorder is added in the form of a random static
transverse field, the average transition probability
for a low-energy
initial state acquires a continuous lineshape. 
Although the dominant delta-like peak found in the ordered case is now absent,
the distribution is still sharply peaked around one of the transitions present
before.
Therefore we conclude that in this case disorder does not
obliterate the occurrence of sharp features in the transition probability
distribution that could be important
for the observability of macroscopic quantum coherence. 
On the other hand, disorder may suppress the backward cascade
effect occurring in the ordered case when the initial state is a highly
excited state. In this case at the end of the time evolution,
the original wavepacket is spread essentially
over all the eigenstates of the system.

The paper is organized as follows.
In Sec.~\ref{radom_spin_ham}
we introduce a giant spin model describing a transition-metal
nanoparticle in the molecular-magnet limit, 
and we illustrate some of its spectral properties.
In Sec.~\ref{dynamics} we
discuss some paradigmatic features of the dynamical evolution of
the collective magnetization under the effect of a 
time-dependent magnetic field.
Disorder-averaged evolution of magnetization from the point of view
of quantum diffusion is discussed in Sec.~\ref{disord_ave}. In
Sec.~\ref{network_model} we interpret our results in the context of
a simplified network model, which offers an intuitive and possibly more
generic picture of the magnetization reversal problem. Our conclusions
are presented in Sec.~\ref{finals}, where we comment on the relevance
of this work for the observation of QTM in ultrasmall ferromagnetic metal
grains.

\section{Giant-spin model for a ferromagnetic metal grain}
\label{radom_spin_ham}

We consider an effective ``spin'' Hamiltonian aimed at modeling
a ferromagnetic transition metal nanoparticle containing on the
order of $N_a \approx 100$ atoms, in such a way that the total
anisotropy energy $KN_a$ is smaller than the single-particle mean-level
spacing. Here $K$ is the bulk anisotropy energy/atom.
In this regime the nanoparticle behaves like a molecular magnet
described by a collective quantum ``spin'' degree
of freedom, $\hat {\bf S}$
\footnote{When the bulk  density of states is used to evaluate $\delta$, 
one finds\cite{ccmPRL03} that 
a transition metal nanoparticle will behave like a molecular magnet
when $N_a < 120$ in Co and $N_a < 750$ in Fe.}.
It is important to emphasize that this collective ``spin'' 
is an effective variable representing coupled quasiparticle
spin {\it and} orbital degrees of freedom of the original
electron system. This coupling
is non-trivial when spin-orbit interaction is included. 
The ``spin'' $S$ can be  
specified by a Berry
curvature Chern number that characterizes the topologically nontrivial
dependence of the many-electron wavefunction on 
magnetization orientation\cite{ccmPRL03}.
The model that we consider has a uniaxial anisotropy term, 
augmented by a ``transverse  
magnetic field'', which can be randomly distributed to mimic the 
combined remnant effect of 
quasiparticle excitations and structural disorder
on the gap structure of the spin 
collective modes\footnote{It is important to point out that
this
random Hamiltonian does
not include the
coupling of the nanoparticle magnetic moment with other types of excitations,
e.g. the ones related to the degrees of freedom of the nuclear spin bath.
Such coupling is likely to be an important source of decoherence.
We will comment on this point in Sec.~\ref{finals}.}.
We will use matrix representations of the (dimensionless) operators 
$(\hat S_x, \hat S_y, \hat S_z)$
in the basis of the eigenstates $\{|S,m>,\  -S \le m \le S\}$
of $(\hat S^{2},\,\hat S_z)$, where $z$ is aligned along the
easy axis. 
The components of the ``transverse field'' are
\begin{equation}
<S,m |{\hat B}_{x}|S,m'> = \delta_{m,m'} r_{x}(m)\Delta_x;\ \
<S,m |{\hat B}_{y}|S,m'> = \delta_{m,m'} r_{y}(m)\Delta_y\;.
\end{equation}
Here $\Delta_x,\,\Delta_y$ are given amplitudes, with dimensions of energy, and
$r_{x}(m),\,r_{y}(m)$ are uniformly distributed
dimensionless random variables having width $1/2$ and zero average.
The resulting random coupling is described by the
symmetrized operators 
\begin{equation}
({\hat B}_{k}\hat S_{k})_{(R)}=(1/2)({\hat B}_{k}\hat S_{k}+\hat S_{k} {\hat B}_{k}),\ \ k=x,y\,,
\label{brandom}
\end{equation}
having matrix elements:
\begin{equation}
<S,m|({\hat B}_{k}\hat S_{k})_{(R)}|S,m- 1> = i^{\epsilon(k)} \Delta_k
(1/2) r'_{k}(m)[(S+m)(S-m+1)]^{1/2}\;,
\end{equation}
where $r'_{k}(m)=[r_{k}(m)+r_{k}(m-1)]/2$ and $ \epsilon(k) = 0,1 (k=x,y)$.
The effective spin Hamiltonian is then:
\begin{equation}
\label{hamiltonian}
\hat H = - B_{z}(t) {\hat S}_{z} - K {\hat S}_{z}^{2} - 
({\hat B}_{x}{\hat S}_{x})_{(R)}- ({\hat B}_{y}{\hat S}_{y})_{(R)}\;,
\end{equation}
where $K$ is an effective anisotropy energy/spin. In Eq.~(\ref{hamiltonian})
we have introduced the coupling to a time-dependent longitudinal
field $B_{z}(t)$, which we will use to manipulate the collective spin spectrum
and induce Landau-Zener type transitions at its avoided-crossing gaps.
It is trivial to verify that $\hat S^2$ commutes
with the operators in Eq.~(\ref{brandom}) and therefore the Hamiltonian
of Eq.~(\ref{hamiltonian}) remains within a given spin multiplet $S$.
The size of the Hilbert space is $2S+1$ and we will label the energy
levels $E(m,t)$ with the discrete index $m$, running from $-S$ to $S$. 
 \begin{figure}
\rotatebox{-90}{\includegraphics[width=8.cm,height=12.cm]{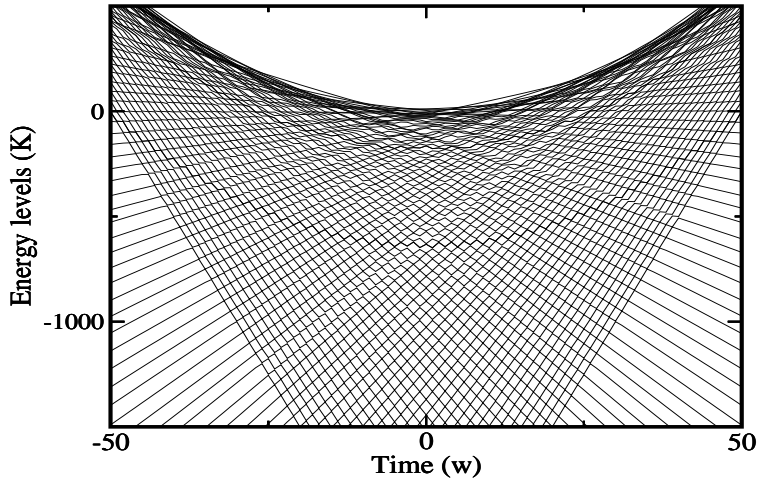}}
 \caption{Energy levels $E(m,t)$ vs. time
of the random spin Hamiltonian defined in Eq.\ref{hamiltonian}, 
with total spin
$S=50$.
The time unit is $w \simeq K/\hbar$, 
where $K$ is the anisotropy energy/spin.} 
 \label{fig1_spettror}
 \end{figure}
We  discuss some properties of the spectrum
when $B_z(t)$ is linearly dependent on time:
$B_z(t) = gt$.
In Fig.~\ref{fig1_spettror} we plot the energy levels $E(m,t)$
as a function of time, for a generic disorder realization of the
Hamiltonian given in Eq.~(\ref{hamiltonian}) when $S=50$.
(The values of the other parameters of the Hamiltonian
are specified in
Sec.~\ref{dynamics}.)
It turns out that
the density of states 
is higher in the upper part of the spectrum;
the avoided crossings there
predominantly involve channels associated with nearby sites along
the $m-$chain.
At lower energies the crossings involve channels
associated with distant sites $m,\;m'$ and
the gaps are accordingly much  smaller, as one can see
by simple perturbative arguments. 
To lowest order the coupling  is the  
product of $l$ nearest neighbor amplitudes $(l = |m'-m|)$.   
These features are clearly visible in Fig.~\ref{fig1_spettror}.
It was argued \cite{ahm_cmc2001ssc} that the peculiar 
diamond-like structure of the spectrum can be a signature of 
ferromagnetic metals.
One can further notice that in the presence of a
constant transverse magnetic field
the nearest neighbor amplitudes
favor backward (forward) motion in the region $m < 0, \;(m > 0)$.
This is due to the angular momentum matrix elements: 
the amplitude of the process $ m \rightarrow m-1$ is larger
than the one relative to $m \rightarrow m+1$ for $m < 0$,
the asymmetry becoming stronger as $m$ approaches the ground state.   
In the classical limit $S >> 1$ one has
a potential barrier separating
the two extremal states $m = \pm S$ \cite{garg0012157}.

\section{Spin Dynamics}
\label{dynamics}
The deterministic evolution of spin systems under the action of a time-dependent
bias has been considered by various authors, mainly in the context of molecular magnets. Different situations have been
considered, ranging from the case of a large, conserved spin multiplet $S$,
to that of a general system of interacting spins 
\cite{garaschi,mina, gara}.
The Landau-Zener theory provides a natural background for this class of problems
\cite{lan, zen,stuck}. The effect of noise on the Landau-Zener
transitions has also been examined at length, in order 
to obtain a better understanding
of the influence of the phonon and spin baths on macroscopic quantum
coherence\cite{saika,posche,sinpro}.
Here, the time evolution of a giant spin with time-independent disorder
is supposed to represent the coherent dynamics of the magnetic moment
of a monodomain ferromagnetic metal nanoparticle in a regime 
characterized by a large energy gap
between the single-particle excitations and the collective ``spin'' modes. 
Our analysis will focus on the effect of quenched disorder.

The wave function 
$|\Psi(t)\rangle $ satisfies
the time-dependent Schr\"odinger equation:
\begin{equation}
\label{schrodinger}
i\hbar{d\,|\Psi(t)\rangle\over d\,t} = 
\hat H(t)|\Psi(t)\rangle
\end{equation}
By expanding $|\Psi(t)\rangle = \sum_m c_{m}(t) |S,m>$ 
on the basis $\{|S,m\rangle\}$, we obtain the following set of coupled 
differential equations for the coefficients $c_{m}(t)$
\begin{eqnarray}
\label{schrodinger2}
i \hbar {\dot c}_{m}(t)&=& (-gtm -K m^2)c_{m}(t)\nonumber\\
- [(\Delta_x/2)r'_{x}(m)&+&i(\Delta_y/2)r'_{y}(m)][(S+m)(S-m+1)]^{1/2}c_{m-1}(t)\\
- [(\Delta_x/2)r'_{x}(m+1)&+&i(\Delta_y/2)r'_{y}(m+1)][(S-m)(S+m+1)]^{1/2}c_{m+1}(t)
\nonumber\; . 
\end{eqnarray}
The Schr\"odinger equation is integrated over a time interval $-T < t < +T$
such that at its extrema  $t = \pm T$ 
the eigenvalues $E(m,t)$ are well separated: the diagonal
part of the Hamiltonian being linear in $t$, and the
coupling with the transverse field being bounded, the eigenvalues approach 
$-gtm$ for large enough $|t|$.
More specifically, under the condition
$|\Delta_x|, |\Delta_y| \ll S$, 
it is sufficient to fix the value of $T$ as follows:
$T > 4 (K S)/g$.
The  eigenvalues  then identify
isolated channels and the time evolution can be studied
as a  scattering problem.
In the present case, the ground state at  $t= -T$ is $m = -S$
and turns into $m = +S$ at $t= +T$.
One would like to         
integrate   the Schr\"odinger equation
over the interval $(-T\;\;,+T)$,
with $S$ on the order of 50 and 
realistic values of the other parameters, such as an
anisotropy energy/spin on the order of $10^{-4}eV$,
and a velocity of the bias $B_{z}(t)$ on the order of mTesla/sec.
It is readily verified that this   
requires astronomical computation times.
The situation is worse if one  needs  averaging over
a large set of disordered configurations.
We would like to emphasize here that in solving the time dependent 
Schr\"odinger
equation, we don't want to make the frequently used approximation of reducing
the full Hamiltonian into an effective two-level model. While this 
procedure is perhaps
a justified approximation in the case of simpler models describing
molecular magnets
with small spins, the level intricacies present in our spectrum, which are
at the heart of the problem we want to study, make this approximation
completely meaningless.  

As an example, we studied the case of the evolution
from the channel $m = -46$, close to the ground state,  
with $K = 10^{-4} eV$ and
$\Delta_x/\mu_{B} = \Delta_y/\mu_{B} = 0.02$ Tesla,
$\mu_{B}$ being the Bohr  magneton.
In a series of runs
we progressively reduced the sweep velocity down to 1 Tesla per second.
Further examination of this case for, say,
a sweep velocity of $10^{-1}$ Tesla per second
requires a few days of computation time.
At 1 Tesla per second, we 
obtained that the system hops from $m = -46$ to $m = -43$.
The question arises then: what can one expect
at smaller sweep velocities?
Could  one  obtain 
a much larger hopping range at, say, 1 mTesla per second?
The Landau-Zener theory gives a negative answer to this question
and a way out of the computation time problem.
Let $\Delta_{\rm eff}(m,m')$ be the (time-dependent) gap between two
channels $m,\;\;m'$; as long as this gap is reasonably
large, wave packets carrying quantum numbers $m$ and $m'$
practically do not interfere.
In the small transient at the crossing time 
(where the gap reaches a minimum) the two channels
do interfere, and their  
scattering 
is determined by the  Landau-Zener 
matrix $S(m,m')$ 
\cite{pokrovsky0012303}.
As is well known, the transition probability $P(m \to m')$
depends on the adimensional parameter
$\nu(m,m') =[\Delta_{\rm eff}(m,m')]^{2}/(g \hbar|m-m'|)$: 
\begin{equation}
P(m \to m') = 1 - exp[- \pi \nu(m,m')/2].
\end{equation}
Notice that $g$ scales as the square of the gap.
Since, as already noticed, the gap corresponding to an hopping event
of range $l$ scales as the $l$-th power of a
perturbative parameter, it is clear that upon
reducing $g$ by few orders of magnitude one will not detect
hoppings of significantly larger range.
This answers the question we made 
above.
On the other hand,
one can  exploit the scaling 
between the gap 
and the sweep velocity in order to infer the behavior
in the physical region from the results corresponding to
numerically affordable values of the parameters.
In a sequence of runs,
we detected
the transitions $m \rightarrow m+1$ , $m \rightarrow m+2$, $m \rightarrow m+3$,
and so on.
We then compared the  crossing times resulting from time integration 
with the ones
extracted by direct inspection of the spectrum $E(m,t)$, 
and found full agreement.
As the crossing behavior appears to be scale invariant,
one can argue that indeed the dynamics can be described as a 
sequence of L-Z transitions.
This argument works  provided that
the hopping events involve two channels at a time.
In the upper part of the spectrum, in particular in
a region around $t = 0$, 
the channels are always close in energy and tend to hybridize. 
The resulting motion is a sequence of short range hoppings,
as we will discuss in the sequel.

Based on the above considerations, in the remaining part of this 
Section we will
consider a value for the sweeping speed that yields reasonable 
calculation times, although it lies beyond the experimentally meaningful range.
Specifically, we introduce the time unit 
\begin{equation}
w = \hbar /K\; ,
\end{equation}
where the anisotropy energy $K$ will be taken as our energy unit.
The sweeping speed is chosen in such a way that $g\cdot w^{2}/\hbar \approx 1$.
As for the values of the transverse amplitudes,  we will take  
$\Delta_{x} = \Delta_{y} \approx 2K$; an accurate estimate of these 
parameters should
be based on microscopic derivations similar to the one suggested 
in Ref.~[\onlinecite{ccmPRL03}],
which is a task beyond the goal of this work. Our choice here is an educated
guess based on the fact that if the gaps of the collective modes of a
ferromagnetic metal nanoparticle 
are due in part to their coupling with quasi-particle excitations,
 in the regime where the mean-level spacing of the latter is larger than
the total anisotropy energy,
the values of the fictitious transverse field 
should lie within a range not smaller than $K$. 
Below we discuss the time evolution of the system
both in the disordered and ordered
case.
In the ordered case the values of the transverse field 
$B^{0}_{x},\; B^{0}_{y}$  have been chosen equal to 
the mean square roots of the random amplitudes:
$ B^{0}_{x} =  (<B^{2}_{x}>)^{1/2} = \Delta_{x} /(2. \cdot 3.^{1/2}) \approx \Delta_{x} \cdot 0.28867$.

\begin{figure}
{\includegraphics[width=10.cm,height=5.cm]{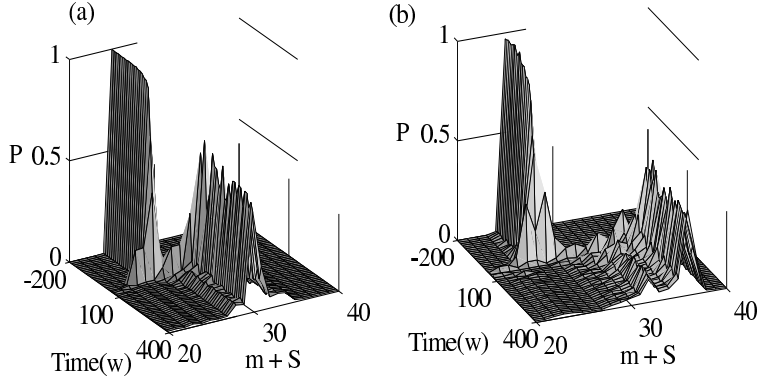}}
 \caption{Probability distribution 
$P(m,t)\equiv |c_{m}(t)|^2= |\langle S,m|\Psi(t)\rangle|^2$ 
as a function of time and the index $m$. The initial wavefunction is prepared
in a state of low energy, $m+S=25$. 
(a) Random case. (b) Deterministic case.}
 \label{fig2_25}
 \end{figure}

We start by examining two paradigmatic behaviors
enabling to reconstruct the generic case.
The first  is found
when the initial state is in the low energy region $m \geq -50$.
Generally the system undergoes a major hopping event toward
a channel $m' > m$, but the hopping range $l = m' -m$, even
if significant, is always far from what one would need for
magnetic inversion.
This is clearly due to the presence of gaps of almost
insurmountable smallness: these, as it is well known,
do occur in the ordered case as well.
It must be recalled that magnetic inversion in molecular 
magnets,
in spite of a smaller spin, is generally observed with the
essential contribution of relaxation and decoherence processes,
which are not taken into account here.
From our  results it appears that in  general the disorder reduces 
the size
of the gaps: more precisely some of the gaps that in the ordered
case are large enough to give rise to hopping, with disorder
are no longer ``seen'', so that the state keeps its quantum
number.  
We also found that as the initial
channel $m,\;(m < 0)$ is closer to the ground state, 
the hopping range becomes shorter.
The reason for this  
has been discussed in Section ~\ref{radom_spin_ham}. 
In Fig.~\ref{fig2_25} we plot the probability distribution
\begin{equation}
P(m,t) \equiv |\langle S,m|\Psi(t)\rangle|^2
\end{equation}
as a function of time and channel index $m$,
for the disordered [Fig.~\ref{fig2_25}(a)] and the ordered 
[Fig.~\ref{fig2_25}(b)] case, respectively.
Only one
disorder realization is considered here.
In both cases the initial wavefunction $|\Psi(t=-T)\rangle$
is a state of sharp $z$-component of the collective spin,
taken in the low-energy part of the spectrum (eigenvalue index $m=-25$),
where the magnetization $\langle \hat S_z \rangle$
is large and negative. 
As shown in Fig.~\ref{fig2_25}(a), in the disordered case
the system undergoes a single 
major hopping event at large positive times.
The result is a rather small magnetization shift
$ \Delta \hat S_{z} = \langle \hat S_{z}(+ T)\rangle - 
\langle\hat S_{z}(- T)\rangle$,
smaller than what one observes in the deterministic case,
shown in Fig.~\ref{fig2_25}(b).
 \begin{figure}
{ \includegraphics[width=10.cm,height=5.cm]{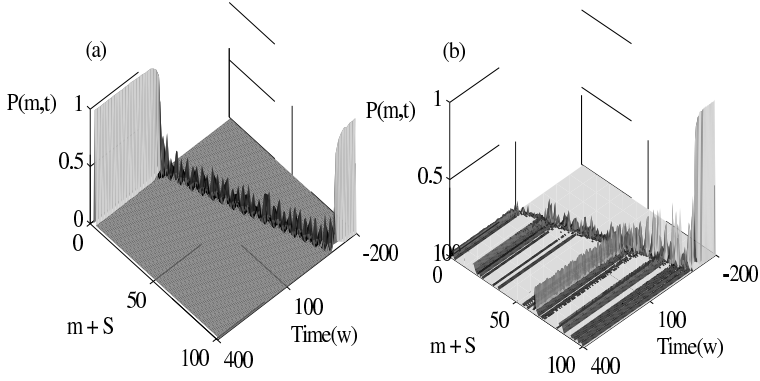}}
 \caption{The same as in Fig.~\ref{fig2_25} but for an initial
wavefunction prepared in an excited state of high energy, $m + S=100$.
(a) Deterministic case. (b) Random case.}
 \label{fig3_100}
 \end{figure}

 \begin{figure}
\rotatebox{-90}{\includegraphics[width=8.cm,height=12.cm]{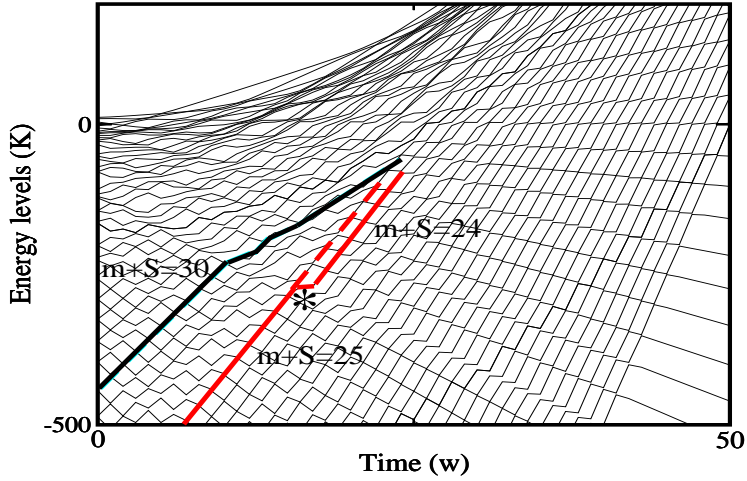}}
 \caption{Energy levels vs. time as in Fig.~\ref{fig1_spettror}. The spectrum
as been blown up in the region corresponding to the 
initial phases of the quantum tunneling
described in Fig.~[\ref{fig2_25}](a).
The gray (red in color version)
bold line represents the time evolution of the energy of
a quantum state initially
prepared in $m+S=25$. At $t\approx 20 w$ (where the asterisk is)
the state first jumps 
into the state  $m+S=24$. Then it starts merging, 
at $t\approx 25 w$, with a narrow group
of states centered around the state with $m+S=30$
(black bold line), which are also moving up and
bending toward the right.
}
 \label{fig1_cross}
 \end{figure}

In order to better see how this transition comes about,
in Fig.\ref{fig1_cross} we plot, as a function of time, the part of the spectrum
blown up around the region where the quantum tunneling for the disordered
case described in 
Fig.~[\ref{fig2_25}](a) occurs. 
The gray (red in color version)
 bold line moving up-right
represents the time evolution of the energy of the initial state,
labeled by $m+S= 25$. As shown in the figure, at time $t\approx 20 w$,
the level starts to encounter large gaps. The first transition is a jump down
into a state of lower $m$; at $t\approx 25 w$, 
the level starts merging with 
a group of states sharply centered around level $m + S = 30$
(represented by the black bold line), which are
also moving up and bending toward the right.
If one starts at higher energies, yet still in the low-energy region
$(\langle \hat S_{z}\rangle < 0)$,  
another mechanism
comes into play, associated with the second of the two announced 
behaviors, which
we will call {\it backward cascading}.
This process,
particularly
clear when the initial state is in the high-energy region,
can be understood as follows.
Every couple of channels $m,m'$ is expected to undergo
a crossing
at the approximate  time  $t_{m,m'}=-K(m+m')/g$,
obtained 
by equating the unperturbed energies : $E^{0}(m,t) = E^{0}(m',t)$,
with $E^{0}(m,t) = -gtm -Km^2$.
The exact crossing time, provided that the
couplings $\Delta_x$ and $\Delta_y$ are small enough, will be perturbatively
close to it.
The nearest-neighbor level crossing   
$(m \to m'= m \pm 1)$ occurs  approximately at the time 
$t_{m} \approx (-2 K m)/g$: the very fact that
$t_{m}$ is a decreasing function of $m$ is at the origin
of the backward cascading process.
In fact, if the system hops 
from $m$ to a higher value $m'$, since $t(m') < t(m)$,
it has no longer a chance to undergo a further nearest-neighbor transition.
If on the contrary  $m' < m$, the next hopping time
has still to come, so that backward motion can be iterated.
One can then expect, upon starting from the highest energy state
$m = S$, a ballistic backward motion ending at $m = -S$.
This is in fact found in the deterministic case, 
as shown in Fig.~\ref{fig3_100}(a).
We have complete magnetic inversion, connecting  energy maxima, 
similar to the pendulum kink, with no dispersion.
The hopping terms in the Hamiltonian drive the 
wave packet down from the local maximum along the energy profile,
but the ballistic velocity equals the time variation
of the energy, so that the wave packet stays on the
maximum.
Disorder inhibits the coherent sequence of nearest neighbor
hoppings; portions of the wavepacket are then trapped at 
intermediate channels; the result is a damped 
backward avalanche, undergoing fragmentations along the way.
Accordingly, the final variance is extremely large, and
complete magnetic inversion is frustrated, as shown
in Fig.~\ref{fig3_100}(b).

In conclusion,
when the initial state of the system is in the low-energy
region, the main feature  of its time-evolution is
the quantum tunneling of the collective spin, associated with a small but not
negligible shift of the magnetization.
When the system starts from an highly excited state, the backward cascading 
is the salient event of the dynamics.
The time evolution of the magnetic system when starting
from a generic state 
appears to be a combination of the two behaviors described above.

\section{Disorder-averaged transition probability}
\label{disord_ave}
 \begin{figure}
{ \includegraphics[width=16.cm,height=12.cm]{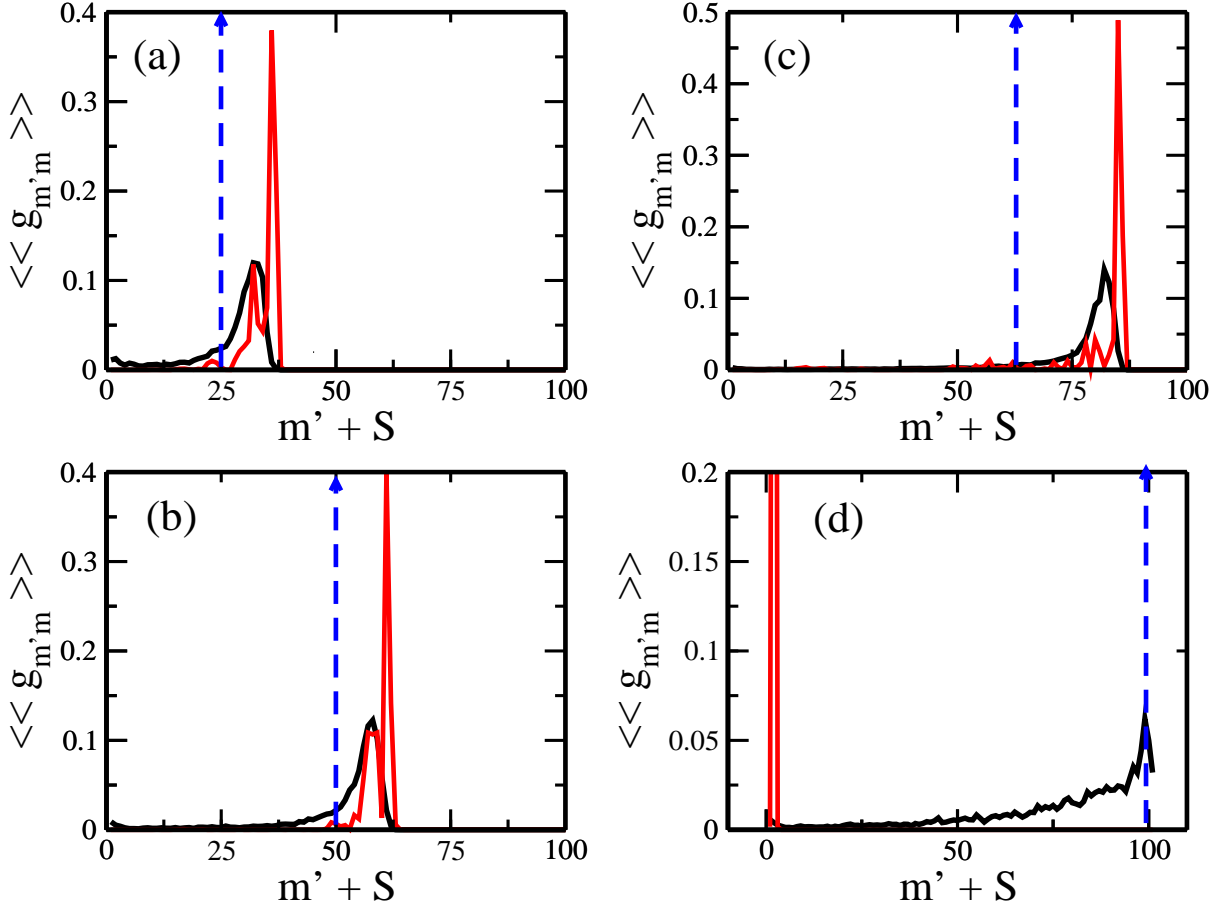}}
 \caption{Disorder-averaged transition probability $<<g_{m',m}>>$ 
(black solid lines). For comparison, the transition probability 
$g_{m',m}$ for
the deterministic case (gray solid lines--red in color version)
is also included.
(a)-(d) represent four different initial conditions.
The vertical dashed line marks the spin quantum number $m$ of the initial
state. Note in (d) the complete ``backward cascade'' behavior,
which turns into a ``cascade with traps'' in the disordered case.}
 \label{fig45_prob}
 \end{figure}
\begin{figure}
\rotatebox{-90}{ \includegraphics[width=8.cm,height=12.cm]{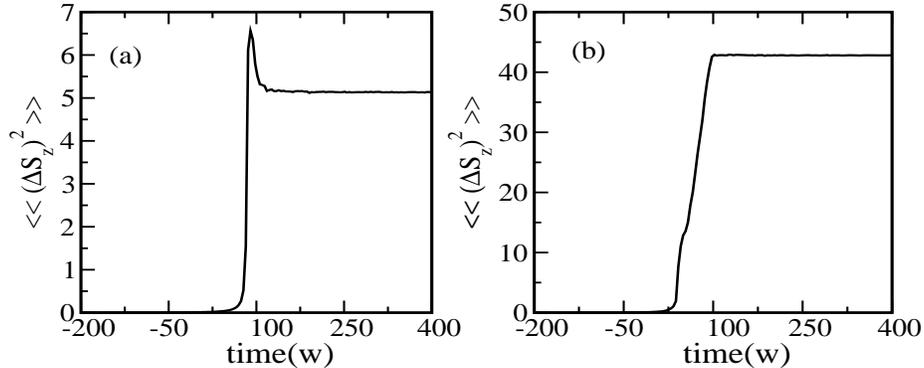}
}
 \caption{Variance of $S_{z}(t)$
averaged over disorder, as a function of time. (a)
Initial state of low energy ($m+S=5$); 
(b) initial state of higher energy ($m+S=25$).}
 \label{fig6_var}
 \end{figure}

So far we illustrated data originating from single samples.
We will now discuss
the disorder-averaged transition probability from the initial state
$m$ to the final state $m'$, defined as
$$
<<g_{m',m}>> = <<|\langle m'|U(+T ; - T)|m\rangle|^{2}>>.
$$
Here, with obvious notation, $U(t' ; t)$ is the evolution operator
from $t$ to $t'$ and the double bracket denotes
the ensemble average.
In Fig.~\ref{fig45_prob} we plot $<<g_{m',m}>>$  (black solid lines)
for the initial conditions (a)-(d) marked by the vertical dashed lines.
The function $g_{m',m}$ for the corresponding 
deterministic case (gray solid lines--red in color version),
with the same initial conditions, is also plotted for comparison.  
The disorder-averaged probability displays a single broad peak at
a value $ m_{max} = m_{max}(m) $, a smooth decay
in the region $m' < m_{max}$ and a very steep decay
on the opposite side of the peak. 
One can notice that $m_{max} > m$, so that it can be identified
with the main hopping event discussed above.
The sensitive asymmetry of the distribution can be explained
in terms of the time ordering of the scattering times:
in fact, after the main hopping event  has taken place, the backward 
cascading  is definitely favored with respect to further forward hoppings.
We also determined the disorder-averaged variance of $S_z$
as a function of time
\begin{equation}
<< (\Delta S_z)^2 >>\, \equiv \, << \langle\Psi(t)|(S_{z} -
< \langle S_{z}\rangle>)^{2}|\Psi(t)\rangle >> 
\end{equation}
for various initial conditions.
This is shown in Fig.~\ref{fig6_var}.

In summary, when starting from low
energy channels (see Fig.~\ref{fig6_var}(a))
there is a peak at the hopping event, then the 
function rapidly decays to a constant value: in fact
during the main transition to $m_{max}$, portions of the wave packet
undergo hoppings of shorter range, and no longer move afterwards.
The picture changes when the initial state has higher energy
(see Fig.~\ref{fig6_var}(b)): then,
following the main hopping event, a cascading process is always present.
This process is a frustrated ballistic motion: the variance 
displays a linear time dependence,
as in quantum diffusion, clearly visible in the time region 
laying between the initial transient
and the final saturation. Notice that the saturation value is
almost one order of magnitude larger than in case (a).
\section{Network of Landau-Zener crossings}
\label{network_model}
\begin{figure}
\rotatebox{-90}{ \includegraphics[width=8.cm,height=10.cm]{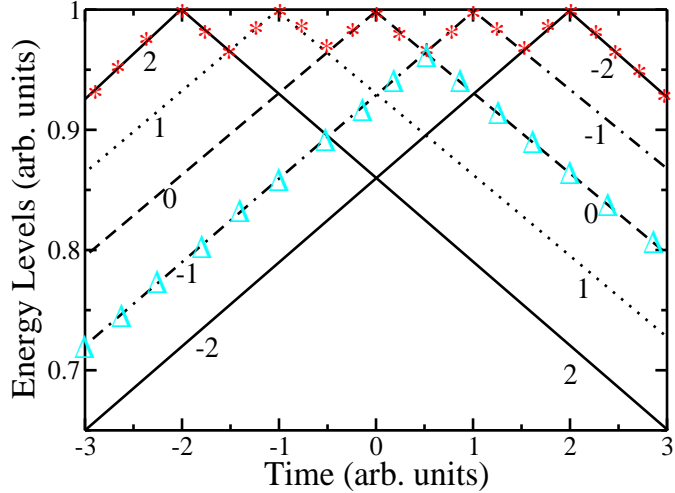}
}
 \caption{Network model in the (t,E) plane, mimicking the time dependence of the level structure
of Fig.\ref{fig1_spettror}. The path marked by the triangles represents
the evolution of the system undergoing quantum tunneling of the magnetization.
The path marked by asterisks represents backward cascade evolution that 
occurs when the system starts from a high-energy state.}
 \label{fig7_network}
 \end{figure}

We now give a qualitative interpretation of our results,
in terms of a simple network model representing the time evolution of the
energy levels. The nodes of the network are associated with the
$m, m'$ crossings. Coherently with this interpretation, we will use
a terminology commonly employed in studying quantum transport
in lattice models, when this   
is viewed as an inter-channel scattering problem described
by the Landauer-B\"uttiker formalism.
Landau-Zener grids were originally introduced in studying the
incoherent mixing of Rydberg manifolds \cite{harpri}. In the approach
suggested below the coherent quantum evolution is instead taken into account.
Due to the presence of the ferromagnetic $S_z^2$ term in the Hamiltonian,
the present network has a peculiar topology, which can be
schematically represented as
a family of parallel lines in the presence of a boundary
acting as a mirror plane. We display this network
in Fig.~\ref{fig1_spettror}, where the energies 
are plotted as a function of time.
All the  
lines are equally oriented with increasing times; 
in the simplest situation the avoided  crossings
involve no more than two lines at a time, 
although, 
as stated in Sec.~\ref{dynamics}, the detailed structure 
of the spectrum is much
more complicated than this and deserves particular care.
A simplified version of this pattern 
is a lattice  where the crossing
times are fixed at their unperturbed values $t_{m,m'}=-K(m+m')/g$,
as depicted in Fig.~\ref{fig7_network}.
Using an optics analogy,
each lattice node acts as a beam splitter, where the 
field amplitude transmission 
corresponds to channel 
conservation $(m \to m)$ and the 
reflection to inter-channel hopping 
$(m \to m')$. 
At the turning points
laying on 
the horizontal line ($E=1$) the channel is conserved.
The time coordinate is oriented from left to right, as in
Fig.~\ref{fig1_spettror}.
The transfer matrix ${\mathcal T}$ for the network of
Fig.~\ref{fig7_network} is the time-ordered product of
single step matrices acting on the space
of wave functions $c_{m}$.
It is convenient
to enumerate the channels with the index 
$n = m + S +1,\; (n=1, N; N = 2 \cdot S +1)$. 
A generic single step matrix, labeled by the ``scattering time''
index $k,\,(k = 3, 2 \cdot N -1)$, is the product of
Landau-Zener  $U(2)$
operators , one for each crossing $(n \rightarrow n')$ occurring
at time $k$.  
The crossings are determined by the equation $n + n' = k;\, (0 < n < n'\le k)$.

Let us discuss the deterministic case first.
One easily realizes that as the distance from the ``mirror' line increases,
the hopping probability decreases; in fact this distance goes as $|m - m'|$,
hence nodes laying far from the ``mirror'' line  involve long 
range hoppings and small gaps, i.e. small Landau-Zener hopping
probabilities.
The network can be separated in two regions, 
respectively dominated by channel conservation
(i.e.localization) 
and by hopping (i.e. delocalization).
The boundary between the two regions, i.e. the ideal
curve where the hopping probability is $1/2$, 
is approximately parallel to the ''mirror'' line.
It is  natural 
to call this boundary
``mobility curve'', although it is not a mobility
edge in the usual sense.  
If, e.g., the particle is initially
in the upper energy level,
it never leaves the
delocalized region: its most probable path is a sequence 
of nearest neighbor hoppings, ending in the final upper level.
This explains the ballistic backward cascade.
This type of time evolution is represented schematically 
on Fig.~\ref{fig7_network} by the path
labeled by asterisks.
When starting from a low energy state,
the particle first propagates in the 
localized region; at some time it will cross the
mobility curve and only at this point it will start hopping, giving rise
typically to one quantum tunneling event only. This second time evolution
is marked on Fig.~\ref{fig7_network} by the path of triangles.
The above analysis can be extended to the disordered case.
One can expect that the 
boundary between localized and delocalized regions
has then a rather intricate shape;
furthermore, since disorder on average lowers the number of 
``large gaps'' the delocalized region  accordingly reduces
its size.
The backwards ballistic motion described above is possible 
provided that
a delocalized strip of almost constant width exists.
If the mobility ``curve'' undergoes fluctuations,
and on average approaches the mirror line,
at the narrowings of  the delocalized strip some portions 
of the wave packet
must enter the localized region.
The result is a frustrated motion, where various portions
of the wave packet get trapped at intermediate states.
If  the particle starts evolving from a
low energy channel, its representative line 
will reach the mobility curve at a later 
time as compared with  the deterministic case: this 
also implies a smaller range $|m - m'|$ of
the main hopping event $m \to m',\, (m' > m)$.
The plots of the transition probability, exhibited in 
Fig.~\ref{fig45_prob}\ 
are consistent with this description; qualitatively the peaks
of these plots identify the average location of the mobility boundary.
\section{Conclusions}
\label{finals}
In this article we have investigated the occurrence of Landau-Zener macroscopic
quantum tunneling of the magnetization in a giant-spin model ($S=50$ )
in the presence of random anisotropy.
The model is intended to provide a phenomenological 
description of the low-energy
spin dynamics of a ultra-small ferromagnetic metal nanograin, in a regime
where the quasi-particle mean level spacing is larger than the total 
anisotropy energy and the metal grain behaves like a molecular
magnet. 
We have focused, in particular, on the effects of the disorder on the gap
structure of the spin collective modes. The microscopic origin of
this randomness is ultimately related to
the non-trivial physics of the itinerant
quasiparticles of the underlaying electronic system.

We find that the time evolution of the system under the
action of a Landau-Zener time-dependent magnetic field depends on the interplay
between disorder and initial conditions. For a disorder-free model
starting from a low-energy state, there is one main coherent
quantum transition event, with a non-negligible shift in the magnetization.
In correspondence of this transition the occupation amplitude
of the original state is essentially totally depleted.
The final (large-time) transition probability distribution is characterized by
a few discrete peaks, with one dominant contribution. 
Disorder does not
obliterate these signatures of macroscopic quantum coherence:
the disorder-averaged transition probability distribution is smooth, sharply peaked and strongly asymmetric.
The resulting shift in magnetization, at a sweep velocity of
the order of mTesla/sec, is estimated in a few  percents
of the spin S.
When the system is initially prepared in the high-energy excited state,
it is subject to multiple tunneling giving rise to
a ballistic motion ending in a final state with
complete magnetization reversal.
This curious coherent time evolution is made possible by
the hopping probability being equal to one during the whole process. 
On the other hand, when disorder is present, at each step the wave packet 
finds a non zero probability of being trapped: as a result
the amplitude of the ballistic wave packet gets damped along the way.

Our results provide some indications about the observability of macroscopic
quantum coherence in ferromagnetic nanoparticles containing approximately
100 atoms. 
Specifically the sharp features in the transition probability,
which are robust against disorder,
can perhaps be observed in detailed Landau-Zener 
SQUID magnetometry experiments on single particles or ensemble of particles,
similar to the ones performed by
Wernsdorfer and colleagues\cite{wernsdorfer97,wernsdorfer2001}.

The scenario presented above excludes macroscopic magnetic
inversion at affordably slow sweep velocities, as well as
hysteresis;
 by considering  decoherence  i.e. the
decay of the non diagonal elements of the density matrix,
and dissipation, i.e. the damping of its diagonal part\cite{leuloss,dobkatz},
these effects should obviously come into play. 
In the analysis done  
in Ref. ~\cite{BS1,BS2},
where the L-Z theory with one level crossing has been generalized to include
decoherence and dissipation,
it was found that the decoherence  does
not completely  destroy the quantum nature of the evolution.
The main features of the Landau-Zener model are thus preserved
in the pure decoherence (non dissipative) case and
survive to a  limited amount of dissipation.
For the large-spin system considered in this paper
the Hamiltonian dynamics
can be described in terms of a sequence of L-Z level
crossings. An important remaining question which we have not investigated 
here is whether or not this picture survives
in an open system. 
The problem of the non-Hamiltonian spin dynamics, already addressed in the
in the past for small quantum spins (see below), is still
the object of intense investigation.

A possible source of dephasing and dissipation is the coupling of the
nanoparticle magnetic moment to a particle-hole continuum, such as a
metallic substrate
and the electron system of the particle itself. As we have explained above,
the latter
should not be important in the molecular magnet regime that we have
considered in this work. The coupling to the
substrate, on the other hand, can be controlled by changing the thickness
of the insulating barrier between the nanoparticle and substrate. 

Another cause of decoherence comes from the coupling to the
phonon bath.
Interestingly enough, the possibility of controlling part
of phonon-induced phenomena has been demonstrated in the case
of the low-spin molecular system $V_{15}$ \cite{chiower}

At low temperatures 
the dominant cause of 
decoherence arises from the unavoidable coupling to nuclear spins.
The central spin model considered 
in Ref.~ \onlinecite{prosta}, where
the microsystem is a single $S = 1/2$ spin, is the best known
description of the spin bath effects;
within that theory, deviations from the Landau-Zener 
behavior  have been  pointed out \cite{sinpro}.
An analysis of spin bath effects on metallic ferromagnetic
nanoparticles goes beyond the scope of the present work.
We recognize that the coupling to nuclear spin baths can considerably
affect the macroscopic quantum coherence studied here.
It is reasonable to assume that this coupling 
can influence to a larger degree the decoherence 
rather than
the dissipation, since the phases of the macroscopic spin
are  sensitive to the nuclear spin precessions, while 
its energies are merely perturbed by the spin bath.

As mentioned in the introduction, so far the only experimental
attempts to detect directly MQT in single ferromagnetic nanoparticles 
has been done
by Wernsdorfer {\it et al.}\cite{wernsdorfer97,wernsdorfer2001}
through magnetization measurements. An idea that we would
like to propose here is to search for evidence of MQT in transport experiments
on magnetic SETs
similar to the ones performed 
by D. Ralph's group\cite{gueron1999,deshmukh2001}, 
but in the presence
of a L-Z time-dependent field\footnote{While we were completing this
work we became aware of a recent paper\cite{QTM_tr_prl04}, where 
a similar suggestion
for the study of MQT in {\it molecular magnets} was put forward.}.
In SET experiments with a magnetic nanoparticle as the central island,
the coupling between
the nanoparticle electronic states and its magnetic moment causes
abrupt changes in the energy of 
conductance resonances at the classical switching field.
This effect was investigated theoretically 
in recent papers\cite{kleff2001prb, ac_cmc_ahm2002}.
An interesting question to ask is how the coherent QTM between
two degenerate quantum states affects the conductance. From the
analysis carried out in this work, we know that  for a large spin
($S\approx 100$) MQT might be observable only at fields close to the
classical switching field. Thus SET transport experiments should
give us a clear landmark of the surroundings of where MQT should be
looked for. How macroscopic quantum coherence would affect the
tunneling resonances is however not obvious and deserves to be further
investigated. Conversely, if MQT in single magnetic particles
could be detected by means of ordinary magnetization measurements,
one very interesting question is to what extent current flow 
will influence dephasing of the
magnetic macroscopic quantum coherence. 
Work along these lines is 
presently underway\footnote{D. Ralph, private communication.}.
As for the coupling of the
nanoparticle to a metallic substrate, the level of decoherence
and dissipation coming from the tunneling current
might be controlled by changing the tunnel barriers of the SET.
\section{Acknowledgments}
We would like to thank W. Wernsdorfer, D. Ralph and A. H. MacDonald
for interesting
discussions.
This work was supported in part by the Swedish Research Council
under Grant No:621-2001-2357 and by the Faculty of Natural Science
of Kalmar University.
Support from the Office of Naval Research under Grant N00014-02-1-0813
is also gratefully acknowledged.
%

\end{document}